\documentclass [11pt]{article}

\title{Anticipating Moves to Prevent Botnet Generated DDoS Flooding Attacks}
\author{Michele Nogueira\footnote{{\bf Notice:} This work has been submitted to the IEEE for possible publication. Copyright may be transferred without notice, after which this version may no longer be accessible.} \\ (michele@inf.ufpr.br)
             }

\begin{document}
\maketitle

\begin{abstract}
Volumetric Distributed Denial of Service (DDoS) attacks have been a recurrent issue on the Internet. These attacks generate a flooding of fake network traffic to interfere with targeted servers or network links. Despite many efforts to detect and mitigate them, attackers have played a game always circumventing countermeasures. Today, there is an increase in the number of infected devices, even more with the advent of the Internet of Things and flexible communication technologies. Leveraging device-to-device short range wireless communications and others, infected devices can coordinate sophisticated botnets, which can be employed to intensify DDoS attacks. The new generation of botnets is even harder to detect because of their adaptive and dynamic behavior yielded by infected mobile portable devices. Additionally, because there can be a large number of geographically distributed devices, botnets increase DDoS traffic significantly. In face of their new behavior and the increasing volume of DDoS traffic, novel and intelligent-driven approaches are required. Specifically, we advocate for {\em anticipating} trends of DDoS attacks in the early stages as much as possible. This work provides an overview of approaches that can be employed to anticipate trends of DDoS attacks generated by botnets in their early stages and brings an insightful discussion about the advantages of each kind of approach and open issues.
\\
\end{abstract}


\section{Introduction}

Volumetric Distributed Denial of Service (DDoS) attacks are still among the top five most compromising threats for the Cyberspace. Traditional DDoS attacks can reach a volume of hundreds of Gigabits per second, generating up to 56 million packets per second~\cite{Man:2015, Ros:2014,Zar:2013}. However, attackers are always enhancing their techniques and taking advantage of the new technologies, such as rapid wireless communication infrastructure, cloud computing and device portability~\cite{Ana:2016, Zha:2012}. More and more, a large amount of infected devices (also known as `bots') coordinate among themselves the generation of thousands of requests against targeted servers or network links~\cite{Far:2016,Fer:2016}. Reports from the last DDoS attacks against the French hosting giant OVH and the Dyn DNS companies show the use of a huge amount of infected devices, including those from the Internet of Things (IoT), such as CCTV cameras, smart DVR recorders and others.  

In addition to the traditional network of bots (botnets), that rely mainly on infected personal computers, sophisticated botnets comprise also of infected portable and mobile devices. These new generation of botnets leverage the advantages of the different communication technologies, such as cellular and device-to-device communications, masking malicious code propagation and facilitating the exploitation of software vulnerabilities and making cheaper to attackers to launch DDoS attacks. Due to the new possibilities of devices portability and the distributed communication between bots, the sophisticated botnets present a dynamic and adaptive behavior, that may circumvent existing defense measures, such as firewalls. Further, because there can be a large number of geographically distributed devices, the new generation of botnets can increase DDoS traffic significantly. The reports of the last DDoS attacks from October 2016 show a volume of traffic around Terabits generated through mobile and portable devices.    

Recent works addressing botnets follow existing approaches designed against traditional botnets. Also, in general, they have focused on assessing malicious code propagation, on botnet detection or DDoS attack mitigation~\cite{Aje:2010,Gel:2016,Lu:2016,Tra:2009,Yu:2015}. However, fighting against the new generation of botnets may require novel, still complementary, and intelligent-driven approaches in order to cope proactively with the new dynamics produced by mobile and portable infected devices. Further, existing defenses may not efficiently mitigate DDoS attacks in progress due to the huge volume of geographically distributed communication requests. 

In face of these limitations, how would be the approaches and techniques to cope with the new generation of botnet-generated DDoS flooding attacks?
We argue in favor of a new perspective that can {\em anticipate} DDoS flooding attacks trends, as much as possible, in their early stages. Hence, this work overviews and reinforces the main techniques that have been contributed towards the DDoS flooding attack trends anticipation, promoting an insightful discussion about them and their application in the context of botnet-generated DDoS flooding attacks. Further, the techniques are conceptually compared considering their advantages and finally the open issues are emphasized, leading to more advances in an intelligent-driven proactive security posture.

\section{New Generation of Botnets}

Botnet refers to a set of malware-infected devices built to perform tasks over the command of its human creator (botmaster). Botnets are used to disrupt the operation of services or extract lucrative gains from unsuspected users. A basic scenario comprises of a single or multiple botmasters, bots, the command and control (C\&C) architecture and its communication protocol. 

Botmasters are essentially human attackers who define all the intelligence inside the botnet, e.g. the attack goal, the target, and the way to infect and coordinate bots. Botmasters rely on a set of predefined commands to indicate the tasks required from bots, such as launching a DDoS attack against a specific target. Today we observe two main classes of botnets, traditional ones and mobile botnets. The differential aspect between them lies in the way bots may communicate to receive updated commands from the botmaster. In addition to traditional wired communication channels, mobile botnets take advantage of mobile network infrastructure to communicate. For instance, certain mobile botnets have the support of SMS transfered over the cellular network infrastructure, others use WiFi or a hybrid approach, mixing device-to-device communication technologies and infrastructured ones, without mentioning the possibility to use the traditional wired channels.

From the C\&C architectural perspective, botnets leverage the classification: $i)$ centralized, $ii)$ distributed or $iii)$ hybrid. In the centralized one, bots need to contact a central server who will translate the goals and actions required by the botmaster into commands to the bots. In a distributed architecture, all bots act simultaneously as servers and clients (i.e., peers) in terms of requesting or indicating actions required by the botmaster. Finally, a hybrid botnet combines the two previous architectures in which some bots are chosen by the botmaster to act as leaders and indicate actions to the other bots.

The new generation of botnets tends to take advantage of the advances in computing and communication technologies~\cite{Ana:2016}. Following a hybrid botnet architecture, bots can quickly adapt the employed communication technology, as well as the communication protocol, policies and target. Bots can mix the advantages of cloud computing with mobile devices' network penetration and high geographical distribution. The possibility to use different communication technologies in order to coordinate the botnet and launch the attack is also an advantage in order to circumvent defenses~\cite{Ana:2016}. The use of learning techniques to understand the network behavior and vulnerabilities helps botmasters in how to configure the botnets, producing more efficient attacks. All these characteristics make the new generation of botnets each more challenging to detect and prevent, requiring then approaches that can cope with the intelligence and heterogeneity, in terms of communication and computing technologies, employed by botmasters.

\section{Anticipating DDoS Attacks Trends}

Given the unpredictable moves of the botmasters and the increasing sophistication of botnets, it becomes harder to rely exclusively on attack detection, measurement and mitigation. In face of the new dynamics produced by the botnets, how would we design new approaches against them? We advocate and reinforce in this article the necessity of creating proactive and intelligent-driven approaches, complementary to the existing ones, against actual and potential threats from botnets and DDoS attacks. It is worth to design solutions towards the creation of adaptive and proactive approaches, that can not only be efficient against the currently known DDoS attacks, but also against unexpected (unknown or unclassified) behaviors they may produce in the future. The rationale for this lies in the fact that attackers are always increasing the level of sophistication taking advantage of the technology advances, and do so defenses should pursue the capability of quickly adapt to this. 

We strongly believe that developing a proactive posture to identify trends of botnets coordination and DDoS flooding attacks as early as possible can help in the application of effective preventive and defensive techniques. Warned about the imminence of a possible DDoS attack coordinated by bots, network administrators and providers can proactively act trying to mitigate and protect the server or network from it. Identifying {\em unclassified} but potentially harmful network behavior based on preliminary indicators is an emerging area of research~\cite{Ram:2016} and it has been motivated by the advances in computation and communication technologies and artificial intelligence. Many challenges exist such as defining the set of generic indicators itself, determine how to promote intelligent gathering and addressing uncertain reasoning and information fusion.

This work contributes with an overview and discussion about the existing initiatives towards the detection of botnets and DDoS flooding attacks trends in their early stages. This discussion is led over three groups, {\em indicators}, {\em techniques} and {\em gathering methods}, in order to facilitate the understanding of efforts through attack early detection. Then, the currently employed indicators, techniques and gathering methods to anticipate trends are highlighted. 
            
\subsection{Indicators}

Determining a set of indicators able to point out possible evidence of botnet coordination and the imminence of DDoS attacks is itself a challenge. First, identifying trends of DDoS attacks and of botnet coordination requires the analysis of a diverse amount of data. Second, there are some different types of flooding DDoS attacks which can be categorized from their initiating feature, such as TCP SYN, ICMP, and new other features the attacker may exploit. Hence, it would be desirable to provide generic indicators that could capture DDoS attack trends and assist in their identification even if the variations in the DDoS attack behavior is unknown or unexpected. Those indicators play an important role on the attack anticipation, once they in general are the basis for the employed techniques.

From the literature, the most common observed approach lies in using a set of attacks or network features that can provide a basis for detecting trends of attacks. For instance, DDoS flooding attacks are known by the huge amount of requests and the intense traffic. While the first intends to overload processing capacity, the later intends to overload network links bandwidth by increasing the size of packets in the network, for instance. Hence, static analysis concentrates on the inspection of the network and communication features, e.g., the number
of packets with specific TCP flags set.
A main drawback of this approach consists in the fact that they are in general designed based on specific characteristics of known DDoS attacks. What in face of the speed and sophistication of the unexpected DDoS behavior may require complementary approaches to be more effective. A real example lies in the recent Mirai botnet that generated peaks of 623 Gbps attack traffic employing exceptionally Generic Routing Encapsulation on packets, what changes from the until know behaviors from previous DDoS attacks.

In a complementary direction, a set of generic statistical indicators has been pointed to identify trends of DDoS attacks in~\cite{Nog:2016}. Those statistical indicators, as return rate, autocorrelation, variance of the load pattern and skewness, have been employed in the context of the Internet and they take as basis the dynamic of the network traffic without being specific to a network or attack feature. Those generic indicators consider the dynamics in the state of the network and consider the Internet as a complex adaptive system. The set of indicators must be employed together and, based on the analysis of their joint behavior, it is possible to warn when the network is shifting abruptly and disruptively from one state to another (i.e. an overloaded state). This is the main aspect taken into account and not the match of the current network behavior with a previously known one.

Finally, it is worth to mention that feature selection and classification techniques from data mining have been employed to point out attributes of the network data traffic relevant for the DDoS attack identification. However, again those techniques have been employed over previously known attacks. Maybe, they can be also a good direction towards the definition of generic indicators, being designed for early detection.


\subsection{Techniques}

Prominently, the majority of the works has concentrated efforts in the choice of the employed technique for early identifying DDoS attacks trends. The technique plays a relevant and critical role, once the volume of data can be very high and in order to anticipate trends the data might need to be processed online and into a short period of time. This subsection provides an overview of the employed techniques. 
This section does not intent to survey the techniques, but provide a picture of them. 

A set of techniques tries to have as reference a previously known behavior and based on that, they look for matching the observed behavior with the known one~\cite{Bij:2016,Tab:2016, Xia:2006,Xyl:2014}. They can be classified as {\em pattern matching}, {\em feature matching} and {\em rule algorithms}, and despite claiming to provide early detection of botnet activities or DDoS attacks, they closely follow approaches employed by traditional DDoS attacks detection. In general, they are based on attack pattern and signature using artificial neural networks, data mining, statistical analysis and hybrid techniques. Artificial neural network methods apply machine learning techniques using patterns or signatures that have already been detected in order to predict whether the network traffic resembles those patterns and can be part of a botnet or a DDoS attack. The accuracy of such techniques depends on various factors, e.g., the training dataset which may not contain patterns of a new type of DDoS attack, or the selected features, and then may not be similar to the DDoS attack in progress and the detection may not be possible.

Further, another data mining technique employed for DDoS attack early detection is a combination of classification and association rule mining algorithms in which a classification technique can be applied to develop a learning model for known attack types while the association rule learning algorithm analyzes the traffic and recognizes relationships between the classes identified in the learning model. Methods based on the Suffix Array data structure is employed to detect all repeated patterns in a sequence. The Suffix Array data structure has been employed over fixed width substrings, such as the IP strings of length 12~\cite{Xyl:2014}. In order to perform this all IP strings are converted to to the same length, i.e., 12 by adding leading zeros in octets that do not have length three, i.e., for single digit octets adding two leading zeros while for double digit octets adding one zero.

Despite these techniques can be applied towards the early detection of botnet-generated DDoS flooding attacks, in general, they have still been employed following a traditional perspective of DDoS attack detection, in which they attempt to detect attack using known indications of attack patterns (these can be either signatures or anomalies) instead of using generic preliminary indications. Given the increasing sophistication of botnets and attacks, it is worth to lead for the detection of unclassified, but potentially harmful behavior, based on preliminary indications before
possible damage occurs. The early detection of unknown attacks and botnet behavior go through the establishment of hypotheses and predictions on not yet understood (unclassified) situations based on preliminary indications, being each time more necessary to deal with uncertainty.

Very few works have addressed the problem considering unclassified and uncertain situations. Those techniques work over the uncertainty of the attack behavior and bot's communication. In~\cite{Kor:2016}, a method for fully distributed early detection is presented. The method is founded on the distributed algorithms that colonies of honey bees use to forage efficiently and provide appropriate dynamic detection thresholds for anomalous event patterns. The approach addresses the challenges of the decentralization promoting cooperation between detector devices (sensors) over an arbitrary virtual topology and allowing coordinated adaptation in the thresholds employed to attack detection by a feedback mechanism. When different attack patterns appear, sensors learn by cooperation to sense them.   

Time-delay neural network (TDNN) is another applied technique. It is a kind of neural network that the time factor is hidden inside the signal with implicit representation. All of the factors of a recent pre-state that may influence the output result in current state is treated as input signal. Therefore, the timing relationship of pre-state and pos-state can be mapped by those signals of data structure. TDNNs follow two kind of architecture. One is multilayer perception which belongs to feed-forward architecture. Another is utilized the rationale of error back propagation to learning and mapping the static relationship between input and output parameters. The concept of static comes from the stationary processes which means that the relationship between input and output will not change with timing factor. The reason that the TDNN has been adopted for early detection of DDoS attacks in~\cite{Tsa:2010} lies in trying to identify the communication between botmaster and bots in the preparation phase of the DDoS attack from the features collected in different timing. Therefore, all of the related features are integrated for associative analysis.

Further, the mathematical property of matrix rank is employed to detect DoS attacks~\cite{Par:2009}. This property allows to analyze the quality of the network traffic once it is extremely sensitive to the randomness in the given matrix. Exploiting this property and the fact that, in general, the major modes of DoS involve randomness (or lack thereof) in the IP addresses, researchers have tried to detect anomaly. In some other works, Gaussian process regression has also been employed to model malicious and/or abnormal behaviors, e.g.~\cite{Fad:2011}. The Gaussian process is a Bayesian data modeling technique that fully accounts for uncertainty. Like other Bayesian-based inference approaches, Gaussian processes have a prior and a posterior. Distributions are defined over functions using the Gaussian process, which is used as a prior for Bayesian
inference. This prior can be flexibly obtained from training or observation data. 
Through Gaussian process regression, samples of a function (model) are observed. Given a set of observation points and their corresponding real valued observations, it is possible to compute the posterior distribution of a new point, which is also Gaussian (i.e., with mean and variance functions). By computing the posterior, it is possible to make predictions for unseen test cases. 

\subsection{Gathering methods}

Gathering data to allow the analysis of the attacks is a issue. Starting from the sensors deployment until the definition of the employed techniques to data fusion and aggregation, all those phases are critical. Different efforts have been made in the literature to improve data gathering with the purpose to assist network security, for instance, the proposed HoneyPots and the recent Darknet~\cite{Fac:2016}. However, addressing the new generation of botnets, such as the mobile botnets, intensifies the challenge once the heterogeneity in the networks brings new levels of complexity.

NEMESYS (Enhanced Network Security for Seamless Service Provisioning in the Smart Mobile Ecosystem) is an example of system that collects and analyzes information about attacks considering the core and mobile network~\cite{Gel:2013}. It provides a data collection infrastructure that incorporates virtualized mobile honeypots and honeyclients in order to gather, detect and provide early warning of mobile attacks. By correlating the extracted information with
known attack patterns from wireline networks, authors try to reveal
and identify the possible shift in the way that attackers launch attacks.                                               

\section{Conclusion and Open Issues}

This work reinforced the increasing threat resulted from the new generation of botnet and pointed out efforts leading to a more proactive security posture based on anticipating trends of botnet and DDoS attack activities. Differently from the traditional network of infected devices, which were based essentially on personal computers, the new generation of botnets take advantage of mobile and portable infected devices and advances in wireless communicate to boost the impact of volumetric DDoS flooding attacks.

This article presented existing initiatives in the direction of early detection of botnet and DDoS flooding attacks following three groups: generic indicators, techniques and gathering methods. Each one of those groups finds itself challenges in the direction of the early detection. The definition of generic indicators depends in a huge set of features and in the understanding of attacks activities as well as the anticipation of unexpected attacks behavior.

Regarding to the employed techniques, the majority of them seems to follow traditional approaches applied for attack detection, even if the focus lies in a early detection. They have demonstrated to focus on pattern/feature/signature matching, being founded on machine learning techniques, artificial intelligence methods and others that not necessary can work efficiently against unclassified or uncertain situations, such as a brand new attack behavior. Few recent works demonstrated to be worried about the uncertainty of attacks behavior. Those works search to applying techniques in direction to a more intelligent-driven security. For instance, the initiatives inspired by Biology that can learn from new patterns and adapt the defense. However, some questions are still open regarding to the employed techniques. For example, how early those techniques addressing uncertainty can detect botnets and DDoS activities? How much data the technique expect as input to have an accurate early detection? And what about false positives and false negatives? How to measure them, considering the uncertain factor? 

Concerning to the data gathering, it is acknowledged as a very important phase for the early detection. Collected data are the basis for analyzing and predicting trends of botnet coordination and DDoS attacks. A work presented some initiatives in using virtual HoneyPots considering the context and behavior expected for the new generation of botnets. Also, some works, such as those associated to the NEMESYS architecture, mention the use of cognitive packets in order to promote a more intelligent data gathering. However, from this author perspective, many issues are still open or unclear. For instance, where should we exactly deploy the sensors? How should we aggregate and provide the data fusion considering that we should work with uncertain situations? Are the existing techniques enough to handle attacks uncertainty?

Finally, this author concludes that despite early detection systems have been mentioned in the literature in a very discrete way for more or less one decade, it seems that the advances in computing and also artificial intelligence have motivated authors to revisit ideas for anticipating trends of botnet and DDoS activities. In the last year, we observe an increase in the number of published works addressing the DDoS issue following this perspective. However, many effort is still necessary in this direction, passing by investigations to determine generic indicators, define better use of artificial intelligence techniques and the complex problem of data gathering and fusion. Open questions exist, and do so opportunities for research.  

\bibliographystyle{IEEEannot}
\bibliography{refs}

\begin{thebibliography}{10}
\providecommand{\url}[1]{#1}
\csname url@rmstyle\endcsname
\providecommand{\newblock}{\relax}
\providecommand{\bibinfo}[2]{#2}
\providecommand\BIBentrySTDinterwordspacing{\spaceskip=0pt\relax}
\providecommand\BIBentryALTinterwordstretchfactor{4}
\providecommand\BIBentryALTinterwordspacing{\spaceskip=\fontdimen2\font plus
\BIBentryALTinterwordstretchfactor\fontdimen3\font minus
  \fontdimen4\font\relax}
\providecommand\BIBforeignlanguage[2]{{%
\expandafter\ifx\csname l@#1\endcsname\relax
\typeout{** WARNING: IEEEtran.bst: No hyphenation pattern has been}%
\typeout{** loaded for the language `#1'. Using the pattern for}%
\typeout{** the default language instead.}%
\else
\language=\csname l@#1\endcsname
\fi
#2}}

\bibitem{Aje:2010}
M.~Ajelli, R.~L. Cigno, and A.~Montresor, ``Modeling botnets and epidemic
  malware,'' in \emph{IEEE International Conference on Communications}, 2010,
  pp. 1--5.


\bibitem{Ana:2016}
M.~Anagnostopoulos, G.~Kambourakis, and S.~Gritzalis, ``New facets of mobile
  botnet: architecture and evaluation,'' \emph{International Journal of
  Information Security}, vol.~15, no.~5, pp. 455--473, 2016.


\bibitem{Bij:2016}
A.~Bijalwan, N.~Chand, E.~S. Pilli, and C.~R. Krishna, ``Botnet analysis using
  ensemble classifier,'' \emph{Perspectives in Science}, vol.~8, pp. 502 --
  504, 2016.


\bibitem{Fac:2016}
C.~Fachkha and M.~Debbabi, ``Darknet as a source of cyber intelligence: Survey,
  taxonomy, and characterization,'' \emph{{IEEE Communications Surveys \&
  Tutorials}}, vol.~18, no.~2, pp. 1197--1227, Second Quarter 2016.


\bibitem{Fad:2011}
Z.~Fadlullah, M.~Fouda, N.~Kato, X.~Shen, and Y.~Nozaki, ``An early warning
  system against malicious activities for smart grid communications,''
  \emph{Network Management of Global Internetworking}, vol.~25, no.~5, pp.
  50--55, 2011.


\bibitem{Far:2016}
P.~Farina, E.~Cambiaso, G.~Papaleo, and M.~Aiello, ``{Are Mobile Botnets a
  Possible Threat? The case of {SlowBot} Net},'' \emph{Computers \& Security},
  vol.~58, pp. 268 -- 283, 2016.


\bibitem{Fer:2016}
E.~Ferrara, O.~Varol, C.~Davis, F.~Menczer, and A.~Flammini, ``{The Rise of
  Social Bots},'' \emph{Communications of the ACM}, vol.~59, no.~7, pp.
  96--104, 2016.


\bibitem{Gel:2013}
E.~Gelenbe, G.~Gorbil, D.~Tzovaras, S.~Liebergeld, D.~Garcia, M.~Baltatu, and
  G.~Lyberopoulos, ``Security for smart mobile networks: The {NEMESYS}
  approach,'' in \emph{International Conference on Privacy and Security in
  Mobile Systems}, 2013, pp. 1--8.


\bibitem{Gel:2016}
E.~Gelenbe, O.~H. Abdelrahman, and G.~Gorbil, ``Detection and mitigation of
  signaling storms in mobile networks,'' in \emph{International Conference on
  Computing, Networking and Communications (ICNC)}, 2016, pp. 1--5.


\bibitem{Kor:2016}
M.~Korczynski, A.~Hamieh, J.~H. Huh, H.~Holm, S.~R. Rajagopalan, and N.~H.
  Fefferman, ``Hive oversight for network intrusion early warning using
  {DIAMoND}: a bee-inspired method for fully distributed cyber defense,''
  \emph{{IEEE Communications Magazine}}, vol.~54, no.~6, pp. 60--67, 2016.


\bibitem{Lu:2016}
Z.~Lu, W.~Wang, and C.~Wang, ``On the evolution and impact of mobile botnets in
  wireless networks,'' \emph{IEEE Transactions on Mobile Computing}, vol.~15,
  no.~9, pp. 2304--2316, 2016.


\bibitem{Man:2015}
S.~Mansfield-Devine, ``{The Growth and Evolution of DDoS},'' \emph{Network
  Security}, no.~10, pp. 13--20, 2015.


\bibitem{Tab:2016}
S.~M.~T. Nezhad, M.~Nazari, and E.~A. Gharavol, ``A novel dos and ddos attacks
  detection algorithm using arima time series model and chaotic system in
  computer networks,'' \emph{IEEE Communications Letters}, vol.~20, no.~4, pp.
  700--703, April 2016.


\bibitem{Nog:2016}
M.~{Nogueira}, ``{Non-Parametric Early Warning Signals from Volumetric DDoS
  Attacks},'' \emph{ArXiv e-prints}, 2016.


\bibitem{Par:2009}
H.~Park, H.~Kim, and H.~Lee, ``Is early warning of an imminent worm epidemic
  possible?'' \emph{Network Management of Global Internetworking}, vol.~23,
  no.~5, pp. 14--20, 2009.


\bibitem{Ram:2016}
A.~A. Ramaki and R.~E. Atani, ``A survey of it early warning systems:
  architectures, challenges, and solutions,'' \emph{Security and Communication
  Networks}, vol.~9, no.~17, pp. 4751--4776, 2016.


\bibitem{Ros:2014}
C.~Rossow, ``{Amplification Hell: Revisiting Network Protocols for DDoS
  Abuse},'' in \emph{Proceedings of the Network and Distributed System Security
  (NDSS) Symposium}, February 2014.


\bibitem{Tra:2009}
\BIBentryALTinterwordspacing
P.~Traynor, M.~Lin, M.~Ongtang, V.~Rao, T.~Jaeger, P.~McDaniel, and
  T.~La~Porta, ``On cellular botnets: Measuring the impact of malicious devices
  on a cellular network core,'' in \emph{Proceedings of the 16th ACM Conference
  on Computer and Communications Security}, ser. CCS '09.\hskip 1em plus 0.5em
  minus 0.4em\relax New York, NY, USA: ACM, 2009, pp. 223--234. [Online].
  Available: \url{http://doi.acm.org/10.1145/1653662.1653690}
\BIBentrySTDinterwordspacing


\bibitem{Tsa:2010}
C.~L. Tsai, A.~Y. Chang, and M.~S. Huang, ``Early warning system for ddos
  attacking based on multilayer deployment of time delay neural network,'' in
  \emph{Sixth International Conference on Intelligent Information Hiding and
  Multimedia Signal Processing}, October 2010, pp. 704--707.


\bibitem{Xia:2006}
B.~Xiao, W.~Chen, and Y.~He, ``A novel approach to detecting ddos attacks at an
  early stage,'' \emph{Journal of Supercomputing}, vol.~36, no.~3, pp.
  235--248, June 2006.


\bibitem{Xyl:2014}
K.~Xylogiannopoulos, P.~Karampelas, and R.~Alhajj, ``{Early DDoS Detection
  Based on Data Mining Techniques},'' in \emph{Proceedings of the IFIP WG 11.2
  International Workshop on Information Security Theory and Practice}, 2014,
  pp. 190--199.


\bibitem{Yu:2015}
S.~Yu, G.~Wang, and W.~Zhou, ``Modeling malicious activities in cyber space,''
  \emph{IEEE Network}, vol.~29, no.~6, pp. 83--87, 2015.


\bibitem{Zar:2013}
S.~T. Zargar, J.~Joshi, and D.~Tipper, ``{A Survey of Defense Mechanisms
  Against Distributed Denial of Service (DDoS) Flooding Attacks},'' \emph{IEEE
  Commun. Surveys \& Tutorials}, vol.~15, no.~4, pp. 2046--2069, 2013.


\bibitem{Zha:2012}
S.~Zhao, P.~P.~C. Lee, J.~C.~S. Lui, X.~Guan, X.~Ma, and J.~Tao, ``Cloud-based
  push-styled mobile botnets: A case study of exploiting the cloud to device
  messaging service,'' in \emph{Proceedings of the 28th Annual Computer
  Security Applications Conference}, 2012, pp. 119--128.


\end{thebibliography}
\end{document}